\documentclass[journal]{IEEEtran}

\usepackage[cmex10]{amsmath}
\usepackage{epsfig,latexsym, amsthm}
\usepackage{amsfonts}
\usepackage{algorithm}
\usepackage{algorithmic}
\usepackage{amsmath}
\usepackage{amsfonts}
\usepackage{epsfig}
\usepackage{url}
\usepackage{hyperref}

\newcommand{\field}[1]{\mathbb{#1}}
\def\R{\field{R}}                                

\newcommand{\Dscr}{{\cal D}}

\newcommand{\Xscr}{{\cal X}}
\newcommand{\Cscr}{{\cal C}}
\newcommand{\Wscr}{{\cal W}}

\newcommand{\Vscr}{{\cal V}}

\newcommand\1{{\bf 1}}
\newcommand\0{{\bf 0}}

\newtheorem{theorem} {Theorem}

\newtheorem{lemma} {Lemma}
\newtheorem{corollary} {Corollary}

\newtheorem{definition} {Definition}

\newcommand{\tends}{\rightarrow}

\newcommand\argmin{\mathop{\mbox{\rm argmin}}\limits}

\begin{document}

\title{Convergence of Min-Sum Message Passing \\
for Quadratic Optimization}

\author{Ciamac~C.~Moallemi,~\IEEEmembership{Member,~IEEE,}
  and~Benjamin~Van~Roy,~\IEEEmembership{Member,~IEEE}
\thanks{C.C. Moallemi is with the Graduate School of Business, Columbia University, New York, NY 10025 USA e-mail: ciamac@gsb.columbia.edu.}
\thanks{B. Van Roy is with the Departments of Management Science \& Engineering and Electrical Engineering, Stanford University, Stanford, CA 94305 USA e-mail: bvr@stanford.edu.}
\thanks{Manuscript received March 14, 2006; revised December 11, 2008.}}

\markboth{IEEE Transactions on Information Theory}%
{Moallemi, Van Roy: Convergence of the Min-Sum Message Passing Algorithm \\
for Quadratic Optimization}

\maketitle

\begin{abstract}
  We establish the convergence of the min-sum message passing
  algorithm for minimization of a quadratic objective function given
  a convex decomposition. Our results also apply to the equivalent
  problem of the convergence of Gaussian belief propagation.
\end{abstract}


\section{Introduction}

\IEEEPARstart{C}{onsider} an optimization problem that is
characterized by a set $\Xscr$ and a hypergraph $(V,\Cscr)$.  There
are $|V|$ decision variables; each is associated with a vertex $i\in
V$ and takes values in a set $\Xscr$. The set $\Cscr$ is a collection
of subsets (or, ``hyperedges'') of the vertex set $V$; each hyperedge
$C\in\Cscr$ is associated with a real-valued ``component function''
(or, ``factor'') $f_C:\ \Xscr^C\rightarrow\R$. The optimization
problem takes the form
\[
\min_{x \in \Xscr^{|V|}} f(x),
\]
where
\[
f(x) = \sum_{C \in \Cscr} f_C(x_C).
\]
Here, $x_C \in \Xscr^{|C|}$ is the vector of variables associated with
vertices in the subset $C$. We refer to an optimization program of
this form as a {\it graphical model}. While this formulation may
seem overly broad---indeed, almost any optimization problem can be
cast in this framework---we are implicitly assuming that the graph is
sparse and that the hyperedges are small.

Over the past few years, there has been significant interest in a
heuristic optimization algorithm for graphical models.  We will call
this algorithm the min-sum message passing algorithm, or the min-sum
algorithm, for short.  This is equivalent to the so-called max-product
algorithm, also known as belief revision, and is closely related to
the sum-product algorithm, also known as belief propagation.  Interest
in such algorithms has to a large extent been triggered by the success
of message passing algorithms for decoding low-density parity-check
codes and turbo codes \cite{Gallager63,Berrou93,Richardson01}.
Message passing algorithms are now used routinely to solve NP-hard
decoding problems in communication systems.  It was a surprise that
this simple and efficient approach offers sufficing solutions.

The majority of literature has been focused on the case where the
set $\Xscr$ is discrete and the resulting optimization
problem is combinatorial in nature. We, however, are interested in the
the case where $\Xscr=\R$ and the optimization problem is
continuous. In particular, many continuous optimization problems that
are traditionally approached using methods of linear programming,
convex programming, etc. also possess graphical structure, with 
objectives defined by sums of component functions. 
We believe the min-sum algorithm leverages this graphical structure in 
a way that can complement traditional optimization algorithms, and
that combining strengths will lead to algorithms that are able to scale to 
larger instances of linear and convex programs.

One continuous case that has been considered in the literature is that
of pairwise quadratic graphical models. Here, the objective function
is a positive definite quadratic function
\begin{equation}\label{eq:quad}
f(x) = \frac{1}{2} x^\top \Gamma x - h^\top x, \quad \Gamma \succ 0.
\end{equation}
This function is decomposed in a pairwise fashion according to an
undirected graph $(V,E)$, so that
\[
f(x) = \sum_{i \in V} f_i(x_i) + \sum_{(i,j) \in E} f_{ij}(x_i,x_j),
\]
where the functions $\{ f_i(\cdot), f_{ij}(\cdot,\cdot)\}$ are quadratic. 
It has been shown that, if the min-sum algorithm converges, it
computes the global minimum of the quadratic
\cite{Weiss01,Rusmevichientong01,Wainwright03}.  The question of
convergence, however, has proved difficult.  Sufficient conditions for
convergence have been established \cite{Weiss01,Rusmevichientong01},
but these conditions are abstract and difficult to verify. Convergence
has also been established for classes of quadratic programs arising in
certain applications \cite{Moallemi06a,Montanari05}.

In recent work, Johnson, et al. \cite{Johnson06,Malioutov06} have
introduced the notion of {\it walk-summability} for pairwise quadratic
graphical models. They establish convergence of the min-sum algorithm for
walk-summable pairwise quadratic graphical models when the particular set of
component functions
\begin{equation}\label{eq:simplef}
f_{ij}(x_i,x_j) = \Gamma_{ij}x_i x_j,\quad\forall\ (i,j)\in E,
\end{equation}
is employed by the algorithm and the algorithm is initialized with
zero-valued messages. Further, they give examples outside this class
for which the min-sum algorithm does not converge.

Note that there may be many ways to decompose a given objective
function into component functions.  The min-sum algorithm takes the
specification of component functions as an input and exhibits
different behavior for different decompositions of the same objective
function. Alternatively, the choice of a decomposition can be seen to
be equivalent to the choice of initial conditions for the min-sum
algorithm \cite{Wainwright03,Wainwright04}.  A limitation of the
convergence result of Johnson, et al. \cite{Johnson06,Malioutov06} is
that it requires use of a particular decomposition of the objective
function of the form \eqref{eq:simplef} and with zero-valued initial
messages.  The analysis presented does not hold in other
situation. For example, the result does not establish convergence of
the min-sum algorithm in the applied context considered in
\cite{Moallemi06a}.

We will study the convergence of the min-sum algorithm given 
a convex decomposition:
\begin{definition}
{\bf (Convex Decomposition)} \\
A {\it convex decomposition} of a quadratic function $f(\cdot)$ is a set of 
quadratic functions $\{ f_i(\cdot), f_{ij}(\cdot) \}$ such that
\[
f(x) = \sum_{i \in V} f_i(x_i) + \sum_{(i,j) \in E} f_{ij}(x_i,x_j),
\]
each function $f_i(\cdot)$ is strictly convex, and each function
$f_{ij}(\cdot,\cdot)$ is convex (although not necessarily strictly
so).
\end{definition}
We will say that a quadratic objective function is {\it convex
  decomposable} if there exists a convex decomposition.  This
condition implies strict convexity of the quadratic objective
function, however, not all strictly convex, quadratic functions are
convex decomposable.

The primary contribution of this paper is in establishing that the
min-sum algorithm converges given {\it any} convex decomposition or
even decompositions that are in some sense ``dominated'' by convex
decompositions. This result can be equivalently restated as a
sufficient condition on the initial messages used in the min-sum
algorithm.  Convergence is established under both synchronous and
asynchronous models of computation. We believe that this is the most
general convergence result available for the min-sum algorithm with a
quadratic objective function.

The walk-summability condition of Johnson, et al. is equivalent to the
existence of a convex decomposition \cite{Malioutov06}. In this way,
our work can be viewed as a generalization of their convergence
results to a broad class of decompositions or initial conditions. This
generalization is of more than purely theoretical interest. The
decentralized and asynchronous settings in which such optimization
algorithms are deployed are typically dynamic. Consider, for example,
a sensor network which seeks to estimate some environmental phenomena
by solving an optimization problem of the form \eqref{eq:quad}. As
sensors are added or removed from the network, the objective function
in \eqref{eq:quad} will change slightly. Reinitializing the
optimization algorithm after each such change would require
synchronization across the entire network and a large delay to allow
the algorithm to converge. If the change in the objective function is
small, it is likely that the change in the optimum of the optimization
problem is small also. Hence, using the current state of the algorithm
(the set of messages) as an initial condition may result in much
quicker convergence. In this way, understanding the robustness of the
min-sum algorithm over different initial conditions is important to
assessing it's practical value.

Beyond this, however, our work suggests path towards understanding the
convergence of the min-sum algorithm in the context of general convex
(i.e., not necessarily quadratic) objective functions. The notion of a
convex decomposition is easily generalized, while it is not clear how
to interpret the walk-summability condition or a decomposition of the
form \eqref{eq:simplef} in the general convex case. In follow-on work
\cite{Moallemi07a}, we have been able to establish such a
generalization and develop conditions for the convergence of the
min-sum algorithm in a broad range of general convex optimization
problems. When specialized to the quadratic case, however, those
results are not as general as the results presented herein.

The optimization of quadratic graphical models can
be stated as a problem of inference in Gaussian graphical models. In
this case, the min-sum algorithm is mathematically equivalent to
sum-product algorithm (belief propagation), or the
max-product algorithm. Our results therefore also apply to Gaussian
belief propagation. However, since Gaussian belief propagation, in
general, computes marginal distributions that have correct means but
incorrect variances, we believe that the optimization perspective is
more appropriate than the inference perspective. As such, we state our
results in the language of optimization.

Finally, note that solution of quadratic programs of the form
\eqref{eq:quad} is equivalent to the solution of the sparse,
symmetric, positive definite linear system $\Gamma x = h$. This is a
well-studied problem with an extensive literature. The important
feature of the min-sum algorithm in this context is that it is
decentralized and totally asynchronous. The comparable algorithms from
the literature fall into the class of classical iterative methods,
such as the Jacobi method or the Gauss-Seidel method
\cite{BertsekasPDP}. In an optimization context, these methods can be
interpreted as local search algorithms, such as gradient descent or
coordinate descent. While these methods are quite robust, they suffer
from a notoriously slow rate of convergence. Our hope is that
message-passing algorithms will provide faster decentralized solutions
to such problems than methods based on local search. In application
contexts where a comparison can be made \cite{Moallemi06a},
preliminary results show that this may indeed be the case.

\section{The Min-Sum Algorithm}

Consider a connected undirected graph with vertices
$V=\{1,\dots,n\}$ and edges $E$.  Let $N(i)$ denote
the set of neighbors of a vertex $i$.  Consider an objective
function $f:\R^n\rightarrow \R$ that decomposes according to pairwise
cliques of $(V,E)$; that is
\begin{equation}\label{eq:ffactor}
f(x) = \sum_{i \in V} f_i(x_i) + \sum_{(i,j) \in E} f_{ij}(x_i,x_j).
\end{equation}

The min-sum algorithm attempts to minimize $f(\cdot)$ by an iterative,
message passing procedure. In particular, at time $t$, each vertex $i$
keeps track of a ``message'' from each neighbor $u\in N(i)$. This
message takes the form of a function $J^{(t)}_{u\rightarrow
  i}:\R\rightarrow\R$. These incoming messages are combined to compute
new outgoing messages for each neighbor. In particular, the message
$J^{(t+1)}_{i\rightarrow j}(\cdot)$ from vertex $i$ to vertex $j\in
N(i)$ evolves according to
\begin{equation}\label{eq:Jupdate}
\begin{split}
\lefteqn{J_{i \rightarrow j}^{(t+1)}(x_j) =} & \\
& \kappa +
\min_{y_i} \left(f_i(y_i) + f_{ij}(y_i,x_j) 
+ \sum_{u \in N(i)\setminus j} J^{(t)}_{u \rightarrow i}(y_i)\right).
\end{split}
\end{equation}
Here, $\kappa$ represents an arbitrary offset term that varies from
message to message.  Only the relative values of the function $J_{i
  \rightarrow j}^{(t+1)}(\cdot)$ matter, so $\kappa$ does not
influence relevant information.  Its purpose is to keep messages
finite.  One approach is to select $\kappa$ so that $J_{i\rightarrow
  j}^{(t+1)}(0) = 0$.  The functions $\{ J^{(0)}_{i\rightarrow
  j}(\cdot) \}$ are initialized arbitrarily; a common choice is to set
$J_{i\rightarrow j}^{(0)}(\cdot) = 0$ for all messages.

At time $t$, each vertex $j$ forms a local objective function
$f^{(t)}_j(\cdot)$ by combining incoming messages according to
\[
f^{(t)}_j(x_j) = \kappa + f_j(x_j) + 
\sum_{i\in N(j)} J^{(t)}_{i\rightarrow j}(x_j).
\]
The vertex then generates a running estimate of the $j$th component of an 
optimal solution to the original problem according to
\[
x^{(t)}_j = \argmin_{y_j} f^{(t)}_j(y_j).
\]
By dynamic programming arguments, it is easy to see that this
procedure converges and is exact given a convex decomposition when 
the graph $(V,E)$ is a tree. We are interested in the case where the graph 
has arbitrary topology.

\subsection{Reparameterizations}

An alternative way to view iterates of the min-sum algorithm is as a series of
``reparameterizations'' of the objective function $f(\cdot)$
\cite{Wainwright03,Wainwright04}. Each reparameterization corresponds
to a different decomposition of the objective function. In particular, 
at each time $t$, we define a function
$f^{(t)}_j:\R\rightarrow\R$, for each vertex $j\in V$, and a function
$f^{(t)}_{ij}:\R^2 \rightarrow \R$, for each edge $(i,j)\in E$, so
that
\[
f(x) = \sum_{i\in V} f^{(t)}_i(x_i) + \sum_{(i,j)\in E} f^{(t)}_{ij}(x_i,x_j).
\]
The functions evolve jointly according to
\[
\begin{split}
f_i^{(t+1)}(x_i) & = \kappa +
f_i^{(t)}(x_i) + \\
& \quad\sum_{j \in N(i)} \min_{x_j} \left(f^{(t)}_j(x_j) + f^{(t)}_{ij}(x_i,x_j)\right),
\end{split}
\]
\begin{equation}\label{eq:Qupdate}
\begin{split}
f_{ij}^{(t+1)}(x_i,x_j) & = \kappa + f_{ij}^{(t)}(x_i,x_j) \\
& \quad - \min_{y_i} \left(f^{(t)}_i(y_i) + f_{ij}^{(t)}(y_i,x_j)\right)
\\
& \quad
- \min_{y_j} \left(f^{(t)}_j(y_j) + f_{ij}^{(t)}(x_i,y_j)\right).
\end{split}
\end{equation}
They are initialized at time $t=0$ according to
\[
f^{(0)}_i(x_i)
=
\kappa
+
f_i(x_i)
+ 
\sum_{j \in N(i)}
J^{(0)}_{j \rightarrow i}(x_i),
\]
\[
\begin{split}
f_{ij}^{(0)}(x_i,x_j)
& =
\kappa
+
f_{ij}(x_i,x_j)
\\
& \quad
-
J^{(0)}_{j \rightarrow i}(x_i)
-
J^{(0)}_{i \rightarrow j}(x_j).
\end{split}
\]
In the common case, where the functions $\{ J^{(0)}_{i\rightarrow
  j}(\cdot)\}$ are all set to zero, the initial component functions
$\{f^{(0)}_i(\cdot), f^{(0)}_{ij}(\cdot,\cdot)\}$ are identical to
$\{f_i(\cdot), f_{ij}(\cdot,\cdot)\}$, modulo constant offsets.  A
running estimate of the $j$th component of an optimal solution to the
original problem is generated according to
\begin{equation}\label{eq:xstarupdate}
x^{(t)}_j = \argmin_{y_j} f^{(t)}_j(y_j).
\end{equation}

The message passing interpretation and the reparameterization
interpretation can be related by
\[
f^{(t)}_j(x_j) = \kappa + f_j(x_j) + 
\sum_{i\in N(j)} J^{(t)}_{i\rightarrow j}(x_j),
\]
\[
f_{ij}^{(t)}(x_i,x_j)
=
\kappa
+
f_{ij}(x_i,x_j)
-
J^{(t)}_{j \rightarrow i}(x_i)
-
J^{(t)}_{i \rightarrow j}(x_j),
\]
\[
\begin{split}
J^{(t+1)}_{i\rightarrow j}(x_j) & =
\kappa
+
J^{(0)}_{i\rightarrow j}(x_j)
\\
&\quad
+
\sum_{s=0}^t
\min_{y_i}
\left(
f^{(s)}_i(y_i) 
+
f_{ij}^{(s)}(y_i,x_j)
\right).
\end{split}
\]
These relations are easily established by induction on $t$.  As they
indicate, the message passing interpretation and the
reparameterization interpretation are completely equivalent in the
sense that convergence of one implies convergence of the other, and
that they compute the same estimates of an optimal solution to the
original optimization problem.

Reparameterizations are more convenient for our purposes for the
following reason: Note that the decomposition \eqref{eq:ffactor} of
the objective $f(\cdot)$ is not unique. Indeed, many alternate
factorizations can be obtained by moving mass between the single
vertex functions $\{ f_i(\cdot)\}$ and the pairwise functions
$\{f_{ij}(\cdot,\cdot)\}$. Since the message passing update
\eqref{eq:Jupdate} depends on the factorization, this would seem to
suggest that the each choice of factorization results in a different
algorithm. However, in the reparameterization interpretation, the
choice of factorization only enters via the initial
conditions. Moreover, it is clear that the choice of factorization is
equivalent to the initial choice of messages $\{ J^{(0)}_{i\rightarrow
  j}(\cdot)\}$.  Our results will identify sufficient conditions on
these choices so that the min-sum algorithm converges.

\section{The Quadratic Case}

We are concerned with the case where the objective function $f$ is
quadratic, i.e.
\[
f(x) = \frac{1}{2}x^\top\Gamma x - h^\top x.
\]
Here, $\Gamma\in \R^{n\times n}$ is a symmetric, positive definite
matrix and $h\in\R^{n}$ is a vector. Since $f$ must decompose
relative to the graph $(V,E)$ according to \eqref{eq:ffactor}, we must
have the non-diagonal entries satisfy $\Gamma_{ij}=0$ if $(i,j)\notin
E$. Without loss of generality, we will assume that $\Gamma_{ij}\neq
0$ for all $(i,j)\in E$ (otherwise, each such edge $(i,j)$ can be
deleted from the graph) and that $\Gamma_{ii}=1$ for all $i\in V$
(otherwise, the variables can be rescaled so that this is true).

Let $\vec{E}\subset V\times V$ be the set of directed edges. That is,
$(i,j)\in E$ iff $\{i,j\}\in \vec{E}$ and $(i,j)\in E$ iff $\{j,i\}\in
\vec{E}$. (We use braces and parentheses to distinguish directed and
undirected edges, respectively.)  Quadratic component functions
$\{f_i(\cdot), f_{ij}(\cdot)\}$ that sum to $f(\cdot)$ can be parameterized
by two vectors of parameters, $\gamma =
(\gamma_{ij})\in\R^{|\vec{E}|}$ and $z=(z_{ij})\in\R^{|\vec{E}|}$,
according to
\[
\begin{split}
f_{ij}(x_i,x_j) 
& = \frac{1}{2} \left(\gamma_{ji} \Gamma_{ij}^2 x_i^2 + 2 \Gamma_{ij} x_i x_j
+ \gamma_{ij} \Gamma_{ij}^2 x_j^2 \right)
\\
&\quad
- z_{ji} x_i - z_{ij} x_j,
\end{split}
\]
\[
f_j(x_j) =
\frac{1}{2} \left(1 - \sum_{i\in N(j)} \Gamma^2_{ij} \gamma_{ij}\right)
x_j^2 - \left(h_j - \sum_{i\in N(j)} z_{ij}\right) x_j.
\]
Given such a representation, we will refer to the components of
$\gamma$ as the quadratic parameters and the components of $z$ as the
linear parameters.

Iterates $\{f_i^{(t)}(\cdot), f_{ij}^{(t)}(\cdot,\cdot)\}$ of the
min-sum algorithm can be represented by quadratic parameters
$\gamma^{(t)}$ and linear parameters $z^{(t)}$.  By explicit
computation of the minimizations involved in the reparameterization
update \eqref{eq:Qupdate}, we can rewrite the update equations in
terms of the parameters $\gamma^{(t)}$ and $z^{(t)}$. In particular,
if $ \sum_{u\in N(i)\setminus j} \Gamma_{ui}^2 \gamma^{(t)}_{ui} < 1$,
then
\begin{equation}\label{eq:gammaupdate}
\gamma^{(t+1)}_{ij} = 
\frac{1}{1 - \sum_{u\in N(i)\setminus j} \Gamma_{ui}^2 \gamma^{(t)}_{ui}},
\end{equation}
\begin{equation}\label{eq:zupdate}
z^{(t+1)}_{ij} = 
\frac{\Gamma_{ij}}{1 - \sum_{u\in N(i)\setminus j} \Gamma_{ui}^2 \gamma^{(t)}_{ui}}
\left( h_i - \sum_{u\in N(i)\setminus j} z^{(t)}_{ui} \right).
\end{equation}
If, on the other hand,
$
\sum_{u\in N(i)\setminus j} \Gamma_{ui}^2 \gamma^{(t)}_{ui} \geq 1,
$
then the minimization
\[
\min_{y_i} f^{(t)}_i(y_i) + f^{(t)}_{ij}(y_i,x_j)
\]
is unbounded and the update equation is ill-posed. Further, the
estimate of the $j$th component of the optimal solution, defined by
\eqref{eq:xstarupdate}, becomes
\begin{equation}\label{eq:xtzupdate}
x^{(t)}_j = 
\frac{1}
{1 - \sum_{i\in N(j)} \Gamma^2_{ij} \gamma^{(t)}_{ij}}
\left(h_j - \sum_{i\in N(j)} z^{(t)}_{ij}\right),
\end{equation}
when $\sum_{i\in N(j)} \Gamma^2_{ij} \gamma^{(t)}_{ij} < 1$, and is
ill-posed otherwise.

We define a generalization to the notion of a convex decomposition.
\begin{definition}
  {\bf (Convex-Dominated Decomposition)}  \\
  A convex-dominated decomposition of a quadratic function $f(\cdot)$
  is a set of quadratic functions $\{f_i(\cdot),f_{ij}(\cdot,\cdot)\}$ that
  form a decomposition of $f(\cdot)$, such that for some convex
  decomposition $\{g_i(\cdot),g_{ij}(\cdot,\cdot)\}$, 
  \[
  g_{ij}(x_i,x_j) - f_{ij}(x_i,x_j)
  \]
  is convex, for all edges $(i,j)\in E$.
\end{definition}
\noindent Note that any convex decomposition is also 
convex-dominated.  

The following theorem is the main result of this paper.
\begin{theorem}
  \label{th:main-convergence}
  {\bf (Quadratic Min-Sum Convergence)} \\
  If $f(\cdot)$ is convex decomposable and
  $\{f^{(0)}_i(\cdot),f^{(0)}_{ij}(\cdot,\cdot)\}$ is a
  convex-dominated decomposition, then the quadratic parameters
  $\gamma^{(t)}$, the linear parameters $z^{(t)}$, and the running
  estimates $x^{(t)}$ converge. Moreover,
\[
  \lim_{t\tends\infty} f(x^{(t)}) = \min_x f(x).
\]
\end{theorem}
\noindent This result is more general than required to capture the
``typical'' situation.  In particular, consider a situation where a
problem formulation gives rise to component functions
$\{f_i(\cdot),f_{ij}(\cdot)\}$ that form a convex decomposition of an
objective function $f$.  Then, initialize the min-sum algorithm with
$\{f^{(0)}_i(\cdot),f^{(0)}_{ij}(\cdot,\cdot)\} =
\{f_i(\cdot),f_{ij}(\cdot,\cdot)\}$.  Since the initial iterate is a
convex decomposition, it certifies that $f(\cdot)$ is convex
decomposable, and it is also a convex-dominated decomposition.

We will prove Theorem~\ref{th:main-convergence} in Section~\ref{se:overall}.
Before doing so, we will study the parameter sequences $\gamma^{(t)}$
and $z^{(t)}$ independently.

\section{Convergence of Quadratic Parameters}\label{se:quad}

The update \eqref{eq:gammaupdate} for the the quadratic parameters
$\gamma^{(t)}$ does not depend on the linear parameters
$z^{(t)}$. Hence, it is natural to study their evolution
independently, as in \cite{Rusmevichientong01,Moallemi06a}. In this
section, we establish existence and uniqueness of a fixed point of the
update \eqref{eq:gammaupdate}. Further, we characterize initial
conditions under which $\gamma^{(t)}$ converges to this fixed point.

Whether or not a decomposition is convex depends on quadratic
parameters but not the linear ones.  Let $\Vscr$ be the set of
quadratic parameters $\gamma\in\R^{|\vec{E}|}$ that correspond to
convex decompositions.

We have the following theorem establishing convergence for the
quadratic parameters. The proof relies on certain monotonicity
properties of the update \eqref{eq:gammaupdate}, and extends the
method developed in \cite{Rusmevichientong01,Moallemi06a}.
\begin{theorem}\label{th:gammaconv}
{\bf (Quadratic Parameter Convergence)}\\
Assume that $f(\cdot)$ is convex decomposable. The system of equations
\[
\gamma_{ij} = 
\frac{1}{1 - \sum_{u\in N(i)\setminus j} \Gamma_{ui}^2 \gamma_{ui}},
\quad\forall\ \{i,j\}\in\vec{E},
\]
has a solution $\gamma^*$ such that
\[
\0 < \gamma^* < v,\quad\forall\ v\in\Vscr.
\]
Moreover, $\gamma^*$ is the unique such solution.

If we initialize the min-sum algorithm so that $\gamma^{(0)} \leq v$,
for some $v\in\Vscr$, then $0 < \gamma^{(t)} < v,$ for all $t > 0$,
and
\[
\lim_{t\rightarrow\infty} 
\gamma^{(t)} = \gamma^*.
\]
\end{theorem}
\begin{IEEEproof}
See Appendix~\ref{ap:gammaconv}.
\end{IEEEproof}

The key condition for the convergence is that the initial
quadratic parameters $\gamma^{(0)}$ must be dominated by those of a 
convex decomposition. Such initial conditions are easy to find, for example
$\gamma^{(0)}=\0$ or $\gamma^{(0)}\in\Vscr$ satisfy this requirement.

Note that we should not expect the algorithm to converge for arbitrary
$\gamma^{(0)}$. For the update \eqref{eq:gammaupdate} to even be
well-defined at time $t$, we require that
\[
\sum_{u\in N(i)\setminus j} \Gamma_{ui}^2 \gamma^{(t)}_{ui} < 1,
\quad\forall\ \{i,j\}\in\vec{E}.
\]
The condition on $\gamma^{(0)}$ in Theorem~\ref{th:gammaconv}
guarantees this at time $t=0$, and the theorem guarantees
that it continue to hold for all $t>0$. Similarly, the computation
\eqref{eq:xtzupdate} of the estimate $x^{(t)}$ requires that
\[
\sum_{i\in N(j)} \Gamma^2_{ij} \gamma^{(t)}_{ij} < 1,
\quad\forall\ j\in V.
\]
The theorem guarantees that this is true for all $t\geq 0$, given
suitable choice of $\gamma^{(0)}$.

\section{Convergence of Linear Parameters}\label{se:lin}

In this section, we will assume that the quadratic parameters
$\gamma^{(t)}$ are set to the fixed point $\gamma^*$, and study the
evolution of the linear parameters $z^{(t)}$. In this case, the update
\eqref{eq:zupdate} for the linear parameters takes the particularly
simple form
\[
z^{(t+1)}_{ij} = 
\gamma^*_{ij}\Gamma_{ij}
\left( h_i - \sum_{u\in N(i)\setminus j} z^{(t)}_{ui} \right).
\]
This linear equation can be written in vector form as
\[
z^{(t+1)} = - D y + A z^{(t)},
\]
where $y\in\R^{|\vec{E}|}$ is a vector with 
\begin{equation}\label{eq:ydef}
y_{ij} = h_i,
\end{equation}
$D\in\R^{|\vec{E}\times\vec{E}|}$ is a diagonal matrix with
\begin{equation}\label{eq:Ddef}
D_{ij,ij} =
-\gamma^*_{ij}\Gamma_{ij},
\end{equation}
and $A\in\R^{|\vec{E}\times\vec{E}|}$ is
a matrix such that
\begin{equation}\label{eq:Adef}
A_{ij,uk} = \begin{cases}
- \gamma^*_{ij} \Gamma_{ij}
& \text{if $(u,i),(i,j)\in E$, $k=i$, $j \neq u$,}
\\
0 & \text{otherwise.}
\end{cases}
\end{equation}
If the spectral radius of $A$ is less than 1, then we have convergence
of $z^{(t)}$ independent of the initial condition $z^{(0)}$ by
\[
\lim_{t\tends\infty} z^{(t)} = - \sum_{t=0}^\infty A^t D y.
\]
We will show that existence of a convex decomposition of $f(\cdot)$ is
a sufficient condition for this to be true.  In order to proceed, we
first introduce the notion of walk-summability.

\subsection{Walk-Summability}

Note that the optimization problem we are considering,
\[
\min_{x}
\frac{1}{2}x^\top\Gamma x - h^\top x,
\]
has the unique solution
\[
x^* = \Gamma^{-1} h.
\]
Define $R = I - \Gamma$, so $R_{ii}=0$ and $R_{ij}=-\Gamma_{ij}$, if
$i\neq j$. If we assume that the matrix $R$ has spectral radius less
than 1, we can express the solution $x^*$ by the infinite
series
\begin{equation}\label{eq:xstardef}
x^* = \sum_{t=0}^\infty R^t h.
\end{equation}
The idea of walk-sums, introduced by Johnson, et al. \cite{Johnson06},
allows us to interpret this solution as a sum of weights of walks on
the graph.

To be precise, define a {\it walk} of length $k$ to be a sequence of
vertices
\[
w = \{w_0, \ldots, w_k\},
\]
such that $(w_i,w_{i+1})\in E$, for all $0\leq i < k$. Given a walk
$w$, we can define a weight by the product
\[
\rho(w) = R_{w_0 w_1} \cdots R_{w_{|w|-1} w_{|w|}}.
\]
(We adopt the convention that $\rho(w)=1$ for walks of length $0$,
which consist of a single vertex.) Given a set of walks $\Wscr$, we
define the weight of the set to be the sum of the weights of the walks
in the set, that is
\[
\rho(\Wscr) = \sum_{w\in \Wscr} \rho(w).
\]
Define $\Wscr_{i\rightarrow j}$ to be the (infinite) set of all walks
from vertex $i$ to vertex $j$. If the quantity
$\rho(\Wscr_{i\rightarrow j})$ was well-defined, examining the
structure of $R$ and \eqref{eq:xstardef}, we would have
\begin{equation}\label{eq:xstarrho}
x^*_{j} = \sum_{i\in V} \rho(\Wscr_{i\rightarrow j}) h_i.
\end{equation}

\begin{definition}
{\bf (Walk-Summability)} \\
  Given a matrix $\Gamma \succ 0$ with $\Gamma_{ii}=1$, define
  $|R|$ by $|R|_{ij}=|[I - \Gamma]_{ij}|$. We say $\Gamma$ is
  walk-summable if the spectral radius of $|R|$ is less than 1.
\end{definition}

Walk-summability of $\Gamma$ guarantees the the function $\rho(\cdot)$
is well-defined even for infinite sets of walks, since in this case,
the series $\sum_{t=0}^\infty R^t$ is absolutely convergent. It is not
difficult to see that existence of a convex decomposition of
$f(\cdot)$ implies walk-summability \cite{Johnson06}. More recent work
\cite{Malioutov06} shows that these two conditions are in fact
equivalent.

We introduce a different weight function $\nu(\cdot)$ defined by
\[
\nu(w) = \gamma^*_{w_0 w_1} R_{w_0 w_1} \cdots 
\gamma^*_{w_{|w|-1} w_{|w|}} R_{w_{|w|-1} w_{|w|}}.
\]
$\nu(\cdot)$ can be extends to sets of walks as before. However, we
interpret this function only over {\it non-backtracking} walks, where
a walk $w$ is non-backtracking if $w_{i-1}\neq w_{i+1}$, for $1 \leq i
< |w|$.  Denote by $\Wscr^{nb}$ the set of non-backtracking walks.
The following combinatorial lemma establishes a correspondence between
$\nu(\cdot)$ on non-backtracking walks and $\rho(\cdot)$.

\begin{lemma}\label{le:nuwalksum}
  Assume that $f(\cdot)$ is convex decomposable.
  For each $w\in\Wscr^{nb}$,
  there exists a set of walks $\Wscr_{w}$, all terminating at the same
  vertex as $w$, such that
\[
\nu(w) = \rho(\Wscr_{w}).
\]
Further, if $w'\in\Wscr^{nb}$ and $w'\neq w$, then $\Wscr_w$ and
$\Wscr_{w'}$ are disjoint.
\end{lemma}
\begin{IEEEproof} See Appendix~\ref{ap:walksum}.
\end{IEEEproof}

The above lemma reveals that $\nu(\cdot)$ is well-defined on infinite
sets of non-backtracking walks. Indeed, if $\Wscr\subset\Wscr^{nb}$,
\begin{equation}\label{eq:nusum}
\sum_{w\in\Wscr} |\nu(w)|
= \sum_{w\in\Wscr} |\rho(\Wscr_{w})|
\leq \sum_{w\in\Wscr} \sum_{u\in\Wscr_{w}} |\rho(u)|,
\end{equation}
and the latter sum is finite since $\Gamma$ is walk-summable.

We can make the correspondence between $\nu(\cdot)$ and $\rho(\cdot)$
stronger with the following lemma.

\begin{lemma}\label{le:nuexact}
  Assume that $f(\cdot)$ is convex decomposable.
  If we define
  $\Wscr^{nb}_{i\rightarrow r}$ to be the set of all non-backtracking
  walks from vertex $i$ to vertex $r$, we have
\[
\rho(\Wscr_{i\rightarrow r}) = \frac{\nu(\Wscr^{nb}_{i \rightarrow r})}
{1 - \sum_{u\in N(r)} R_{ur}^2 \gamma^*_{ur}}.
\]
\end{lemma}
\begin{IEEEproof} See Appendix~\ref{ap:walksum}.
\end{IEEEproof}

\subsection{Spectral Radius of $A$}

Examining the structure of the matrix $A$ from \eqref{eq:Adef}, it is
clear that if $\Wscr^{nb,t}_{uk\rightarrow ij}$ is defined to be the
set of all length $t$ non-backtracking walks $w$ with
$\{w_0,w_1\}=\{u,k\}$ and $\{w_{|w|-1},w_{|w|}\}=\{i,j\}$, then
\[
[A^tD]_{ij,uk} = \nu(\Wscr^{nb,t+1}_{uk\rightarrow ij}).
\]
Thus, if $\Wscr^{nb,1+}_{uk\rightarrow ij}$ is the set of all
non-backtracking walks $w$ of length at least 1 satisfying $\{w_0,w_1\}=\{u,k\}$ and
$\{w_{|w|-1},w_{|w|}\}=\{i,j\}$,
\[
\begin{split}
\sum_{t=0}^\infty [A^tD]_{ij,uk}
& =
\sum_{t=0}^\infty \nu(\Wscr^{nb,t+1}_{uk\rightarrow ij})
=
\nu(\Wscr^{nb,1+}_{uk\rightarrow ij})
\\
&
=
\sum_{w\in\Wscr^{nb,1+}_{uk\rightarrow ij}}
\nu(w).
\end{split}
\]
Lemma~\ref{le:nuwalksum} and \eqref{eq:nusum} assure us that the later
sum must be absolutely convergent. Then, we have established the
following lemma.

\begin{lemma}\label{le:Atconv}
Assume that $f(\cdot)$ is convex decomposable.
The spectral radius of $|A|$ is less than 1.
\end{lemma}

\subsection{Exactness}

From Lemma~\ref{le:Atconv}, we have
\[
z^{(\infty)} = \lim_{t\tends\infty} z^{(t)} = - \sum_{t=0}^\infty A^t D y.
\]
For each vertex $j$, define the quantity
\[
\bar{\Gamma}_j = 
\frac{1}{1 - \sum_{i\in N(j)} \Gamma_{ij}^2 \gamma^*_{ij}}.
\]
In this case, the estimate $x_j^{(t)}$ for each vertex $j$, defined by
\eqref{eq:xstarupdate}, converges to
\[
\begin{split}
x^{(\infty)}_j
&
= 
\bar{\Gamma}_j
\left(
h_j 
- z^{(\infty)}
\right)
\\
&
= 
\bar{\Gamma}_j
\left(
h_j 
+ \sum_{i\in N(j)} \sum_{t=0}^\infty [A^tD y]_{ij}
\right)
\\
&
=
\bar{\Gamma}_j
\left(
h_j + 
\sum_{i \in N(j)}
\sum_{\{u,k\}\in\vec{E}}
\nu(\Wscr^{nb,1+}_{uk\rightarrow ij})
h_u
\right)
\\
&
=
\bar{\Gamma}_j
\left(
h_j + 
\sum_{u\in V}
\nu(\Wscr^{nb,1+}_{u\rightarrow j})
h_u
\right).
\end{split}
\]
Here, we define $\Wscr^{nb,1+}_{u\rightarrow j}$ is the set of
non-backtracking walks of length at least $1$ starting at $u$ and
ending at $j$.  Note that if $u\neq j$, then a non-backtracking walk
from $u$ to $j$ must have length at least 1. Thus,
\[
\nu(\Wscr^{nb}_{u\rightarrow j})=\nu(\Wscr^{nb,1+}_{u\rightarrow j}).
\]
If $u=j$, there is a single non-backtracking walk of length 0 from $j$
to $j$, namely $w=\{j\}$, and $\nu(w)=1$. Thus,
\[
\nu(\Wscr^{nb}_{u\rightarrow j})=1 + \nu(\Wscr^{nb,1+}_{u\rightarrow j}).
\]
Hence,
\[
x^{(\infty)}_j
=
\frac{1}{1 - \sum_{i\in N(j)} \Gamma_{ij}^2 \gamma^*_{ij}}
\sum_{u\in V}
\nu(\Wscr^{nb}_{u\rightarrow j})
h_u.
\]
Comparing with Lemma~\ref{le:nuexact}, and \eqref{eq:xstarrho}, we
have
\[
x^{(\infty)}_j
=
\sum_{u\in V}
\rho(\Wscr_{u\rightarrow j})
h_u
= x^*_j.
\]
Thus, $x^{(\infty)}=x^*$.

Putting together the results in this section, we have the following theorem.

\begin{theorem}\label{th:zconv}
{\bf (Linear Parameter Convergence)}\\
Assume that $f(\cdot)$ is convex decomposable and that
$\gamma^{(0)}=\gamma^*$.  Then, for arbitrary initial conditions
$z^{(0)}$, the linear parameters $z^{(t)}$ converge. Further, the
corresponding estimates $x^{(t)}$ converge to the global optimum
$x^*$.
\end{theorem}

\section{Overall Convergence}
\label{se:overall}

In Section~\ref{se:quad}, we established the convergence
of the quadratic parameters $\gamma^{(t)}$. In Section~\ref{se:lin},
we established the convergence of the linear parameters $z^{(t)}$
assuming the quadratic parameters were set to their fixed point. Here,
we will combine these results in order to prove
Theorem~\ref{th:main-convergence}, which establishes convergence of
the full min-sum algorithm, where the linear parameters evolve jointly
with the quadratic parameters. 

It suffices to establish convergence of the linear parameters
$z^{(t)}$.  Define the matrix $A^{(t)}\in\R^{|\vec{E}\times\vec{E}|}$
by
\[
A^{(t)}_{ij,uk} = \begin{cases}
- \gamma^{(t+1)}_{ij} \Gamma_{ij}
& \text{if $(u,i),(i,j)\in E$, $k=i$, $j \neq u$,}
\\
0 & \text{otherwise.}
\end{cases}
\]
Define the diagonal matrix $D^{(t)}\in\R^{|\vec{E}\times\vec{E}|}$ by
$D^{(t)}_{ij,ij} = -\gamma^{(t+1)}_{ij}\Gamma_{ij}$. Then, the
min-sum update \eqref{eq:zupdate} becomes
\[
z^{(t+1)} = - D^{(t)} y + A^{(t)} z^{(t)},
\]
where $y$ is defined by \eqref{eq:ydef}. From
Theorem~\ref{th:gammaconv}, it is clear that $A^{(t)}\rightarrow A$
and $D^{(t)}\rightarrow D$ (where $A$ and $D$ are defined by
\eqref{eq:Adef} and \eqref{eq:Ddef}, respectively).

From Lemma~\ref{le:Atconv}, the spectral radius of $|A|$ is less than
1. Hence, there is a vector norm $\|\cdot\|$ on $\R^{|\vec{E}|}$ and a
corresponding induced operator norm such that $\| A \| < \alpha$, for
some $\alpha < 1$ \cite{Horn85}. Pick $K_1$ sufficiently large so that
$\| A^{(t)} \| < \alpha$ for all $t \geq K_1$. Then, the series
\[
\sum_{s=0}^\infty \left(A^{(t)}\right)^s
\]
converges for $t \geq K_1$. Set
\[
w^{(t)} = - \sum_{s=0}^\infty \left(A^{(t)}\right)^s D^{(t)} y = -(I - A^{(t)})^{-1} D^{(t)} y,
\]
\[
z^{(\infty)} = -\sum_{s=0}^\infty A^s D y = - (I - A)^{-1} D y.
\]
Then, for $t \geq K_1$,
\[
\begin{split}
\| z^{(t+1)} - z^{(\infty)} \|
& 
\leq \| A^{(t)} (z^{(t)} - w^{(t)}) \| + \| z^{(\infty)} - w^{(t)} \|
\\
& 
\leq \alpha \| z^{(t)} - w^{(t)} \| + \| z^{(\infty)} - w^{(t)} \|
\\
&
\leq \alpha \| z^{(t)} - z^{(\infty)} \| 
+ (1 + \alpha) \| z^{(\infty)} - w^{(t)} \|.
\end{split}
\]
Since $w^{(t)}\rightarrow z^{(\infty)}$, for any $\epsilon > 0$ we can pick $K_2
\geq K_1$ so that if $t > K_2$, $\|w^{(t)} - z^{(\infty)}\| < \epsilon$. Then,
for $t > K_2$,
\[
\| z^{(t+1)} - z^{(\infty)} \|
< \alpha \| z^{(t)} - z^{(\infty)} \| + (1 + \alpha)\epsilon.
\]
Repeating over $t$,
\[
\| z^{(t)} - z^{(\infty)} \|
<
\alpha^{t-K_2} \| z^{(K_2)} - z^{(\infty)} \| + \frac{1+\alpha}{1 - \alpha}\epsilon.
\]
Thus,
\[
\limsup_{t\rightarrow \infty}
\| z^{(t)} - z^{(\infty)} \|
\leq
\frac{1+\alpha}{1 - \alpha}\epsilon.
\]
Since $\epsilon$ is arbitrary, it is clear that $z^{(t)}$ converges to
$z^{(\infty)}$. 

The fact that $x^{(t)}$ converges to $x^*$ follows from the same
argument as in Theorem~\ref{th:zconv}.

\subsection{Asynchronous Convergence}

The work we have presented thus far considers the convergence of a
synchronous variation of the min-sum algorithm. In that case, every
component of each of the parameter vectors $\gamma^{(t)}$ and
$z^{(t)}$ is update at every time step. However, the min-sum algorithm
has a naturally parallel nature and can be applied in distributed
contexts. In such implementations, different processors may be
responsible for updating different components of the parameter
vector. Further, these processors may not be able to communicate at
every time step, and thus may have insufficient information to update
the corresponding components of the parameter vectors. There may not
even be a notion of a shared clock. As such, it is useful to consider
the convergence properties of the min-sum algorithm under an {\it
  asynchronous} model of computation.

In such a model, we assume that a processor associated with vertex $i$
is responsible for updating the parameters $\gamma^{(t)}_{ij}$ and
$z^{(t)}_{ij}$ for each neighbor $j\in N(i)$. We define the $T^{i}$ to
be the set of times at which these parameters are updated. We define
$0 \leq \tau_{ji}(t) \leq t$ to be the last time the processor at
vertex $j$ communicated to the processor at vertex $i$. Then, the
parameters evolve according to
\[
\gamma^{(t+1)}_{ij} = 
\begin{cases}
\frac{1}{1 - \sum_{u\in N(i)\setminus j} \Gamma_{ui}^2 \gamma^{(\tau_{ui}(t))}_{ui}}
& \text{if $t\in T^i$},
\\
\gamma^{(t)}_{ij}
& \text{otherwise,}
\end{cases}
\]
\[
z^{(t+1)}_{ij} = 
\begin{cases}
\frac{\Gamma_{ij}\left( h_i - \sum_{u\in N(i)\setminus j} z^{(\tau_{ui}(t))}_{ui} \right)
}{1 - \sum_{u\in N(i)\setminus j} \Gamma_{ui}^2 \gamma^{(\tau_{ui}(t))}_{ui}}
& \text{if $t\in T^i$,}
\\
z^{(t)}_{ij}
& \text{otherwise,}
\end{cases}
\]
Note that the processor at vertex $i$ is not computing its updates
with the most recent values of the other components of the parameter
vector. It uses the values of components from the last time it
communicated with a particular processor.

We will make the assumption of {\it total asynchronism}
\cite{BertsekasPDP}: we assume that each set $T^i$ is infinite, and
that if $\{t_k\}$ is a sequence in $T^i$ tending to infinity, then
$\lim_{k\rightarrow \infty} \tau_{ij}(t_k) = \infty$, for each
neighbor $j \in N(i)$. This mild assumption guarantees that each
component is updated infinitely often, and that processors eventually
communicate with neighboring processors. It allows for arbitrary
delays in communication, and even the out-of-order arrival of messages
between processors.

We can extend the convergence result of
Theorem~\ref{th:main-convergence} to this setting. The proof is
straightforward given the results we have already established and
standard results on asynchronous algorithms (see \cite{BertsekasPDP},
for example). We will provide an outline here. For the convergence
of the quadratic parameters, note that the synchronous iteration
\eqref{eq:gammaupdate} is a monotone mapping (see Lemma~\ref{le:Fprop}
in Appendix~\ref{ap:gammaconv}). For such monotone mappings,
synchronous convergence implies totally asynchronous convergence by
Proposition~6.2.1 in \cite{BertsekasPDP}. The linear parameter update
equation for the synchronous algorithm has the form
\[
z^{(t+1)} = - D^{(t)} y + A^{(t)} z^{(t)}.
\]
For $t$ sufficiently large, by the convergence of the quadratic
parameters, the matrix $A^{(t)}$ becomes arbitrarily close to
$A$. From Lemma~\ref{le:Atconv}, the matrix $|A|$ has
spectral radius less than one. In this case, by Corollary~2.6.2 in
\cite{BertsekasPDP}, it must correspond to a weighted maximum norm
contraction. Then, one can establish asynchronous convergence of the
linear parameters by appealing again to Proposition~6.2.1 in
\cite{BertsekasPDP}.

\section{Discussion}

The following corollary is a restatement of Theorem~\ref{th:main-convergence}
in terms of message passing updates of the form \eqref{eq:Jupdate}.
\begin{corollary}\label{co:conv}{\bf (Convergence of Message Passing Updates)}\\
  Let $\{g_i(\cdot),g_{ij}(\cdot,\cdot)\}$ be a convex decomposition
  of $f(\cdot)$, and let $\{f_i(\cdot),f_{ij}(\cdot)\}$ be a
  decomposition of $f(\cdot)$ into quadratic functions such that
\begin{equation}\label{eq:convmessage}
g_{ij}(x_i,x_j) + J^{(0)}_{i\rightarrow j}(x_j) + J^{(0)}_{j\rightarrow i}(x_i) - f_{ij}(x_i,x_j)
\end{equation}
is a convex function of $(x_i,x_j)$, for all $(i,j)\in E$.
Then, using the decomposition $\{f_i(\cdot),f_{ij}(\cdot,\cdot)\}$ and
quadratic initial messages $\{J^{(0)}_{i\rightarrow j}(\cdot)\}$, the
running estimates $x^{(t)}$ generated by the min-sum algorithm
converge. Further,
\[
\lim_{t\tends\infty}
f(x^{(t)}) = \min_x f(x).
\]
\end{corollary}

The work of Johnson, et al. \cite{Johnson06} identifies existence of convex
decomposition of the objective as a important condition for such
convergence results and also introduces the notion of
walk-summability. However, the convergence analysis presented there
only establishes a special case of the above corollary, where
\[
f_{ij}(x_i,x_j) = \Gamma_{ij} x_i x_j,
\ \forall\ (i,j)\in E,
\]
\[
J^{(0)}_{i\rightarrow j}(x_j) = 0,
\ \forall\ \{i,j\}\in \vec{E}.
\]
In addition, they present a quadratic program that is not convex
decomposable, and where the min-sum algorithm fails to converge.

The prior work of the current authors in \cite{Moallemi06a} considers a
case that arises in distributed averaging applications. There,
convergence is established when
\[
f_{ij}(x_i,x_j) = \frac{1}{2} \Gamma_{ij} (x_i - x_j)^2,\ \Gamma_{ij}>0,
\ \forall\ (i,j)\in E,
\]
\[
\text{$J^{(0)}_{i\rightarrow j}(\cdot)$ is convex},
\ \forall\ \{i,j\}\in \vec{E},
\]
This is also a special case of Corollary~\ref{co:conv}. The work in
\cite{Moallemi06a} further develops complexity bounds on the rate of
convergence in certain special cases. Study of the rate of convergence
of the min-sum algorithm in more general cases remains an open issue.

Note that the main convexity condition \eqref{eq:convmessage} of
Corollary~\ref{co:conv} can also be interpreted in the context of
general convex objectives. While our analysis is very specific to the
quadratic case, the result may be illuminating in the broader context
of convex programs.

Finally, although every quadratic program can be decomposed over
pairwise cliques, as we assume in this paper, there may also be
decompositions involving higher order cliques. Our analysis does not
apply to that case, and this is an interesting question for future
consideration.

\appendices

\section{Proof of Theorem~\ref{th:gammaconv}}\label{ap:gammaconv}

Define the domain
\[
\Dscr = 
\left\{ \gamma \in \R^{|\vec{E}|}\ \left| \ 
\sum_{u\in N(i)\setminus j} \Gamma_{ui}^2 \gamma_{ui} < 1,
\ \forall\ \{i,j\}\in\vec{E}
\right.
\right\},
\]
and the operator $F:\Dscr \rightarrow \R^{|\vec{E}|}$ by
\[
F_{ij}(\gamma) = 
\frac{1}{1 - \sum_{u\in N(i)\setminus j} \Gamma_{ui}^2 \gamma_{ui}},
\quad\forall\ \{i,j\}\in\vec{E}.
\]
This operator corresponds to a single min-sum update
\eqref{eq:gammaupdate} of the quadratic parameters.  We will first
establish some properties of this operator.

\begin{lemma}\label{le:Fprop}
The following hold:
\newcounter{lcount}
\begin{itemize}{}{}
\item[(i)] The operator $F(\cdot)$ is continuous.
\item[(ii)] The operator $F(\cdot)$ is monotonic. That is, if
  $\gamma,\gamma'\in\Dscr$ and $\gamma \leq \gamma'$, $F(\gamma) \leq
  F(\gamma')$.
\item[(iii)] The operator $F(\cdot)$ is positive. That is, if
  $\gamma\in\Dscr$, $F(\gamma)>\0$.
\item[(iv)] If $v\in\Vscr$ and $\gamma \leq v$,
\[
\alpha F(\gamma) < (\alpha - 1) v + F(v - \alpha(v - \gamma)),
\quad\forall\ \alpha > 1.
\]
\item[(v)] If $v\in\Vscr$, $F(v) < v$.
\end{itemize}
\end{lemma}
\begin{IEEEproof}
Parts~(i)-(iii) follow from the corresponding properties of the function
\[
x\mapsto \frac{1}{1-x},
\]
for $x \in (-\infty,1)$. Part~(v) follows from setting $\gamma=v$ in
Part~(iv).

Part~(iv) remains. For notational convenience, define 
\[
R_{ij}(\gamma) = 1 - \sum_{u\in N(i)\setminus j} \Gamma_{ui}^2 \gamma_{ui},
\]
\[
z = v - \gamma \geq 0.
\]
We have
\[
\begin{split}
\lefteqn{
(\alpha - 1) v_{ij} 
+ 
F_{ij}(v - \alpha (v -\gamma))
-
\alpha F_{ij}(\gamma)
}
&
\\
&
=
(\alpha - 1) v_{ij}
+
\frac{1}{
R_{ij}(v-\alpha z)
}
-
\frac{\alpha}
{R_{ij}(v-z)}
\\
&
=
\frac{
1
}
{
R_{ij}(v - \alpha z)
R_{ij}(v-z)
}
\\
&\quad\ \ 
\times \big\{
(\alpha - 1) v_{ij} 
R_{ij}(v - \alpha z)
R_{ij}(v-z)
\\
&\quad\quad\quad
+
R_{ij}(v-z)
-
\alpha
R_{ij}(v - \alpha z)\big\}
.
\end{split}
\]
Denote the numerator of the last expression by $\Delta$. Since the
denominator is positive, it suffices to show that $\Delta>0$. Define
\[
V_j = 1 - \sum_{i\in N(j)} \Gamma_{ij}^2 v_{ij} > 0,
\]
\[
S_{ij} = \sum_{u\in N(i)\setminus j} \Gamma_{ui}^2 z_{ui} \geq 0.
\]
Note that
\[
R_{ij}(v - \alpha z)
= V_i + \Gamma_{ij}^2 v_{ji} + \alpha S_{ij},
\]
\[
R_{ij}(v - z) = V_i + \Gamma_{ij}^2 v_{ji} + S_{ij}.
\]
Since $v\in\Vscr$, we have $\Gamma_{ij}^2 v_{ij}v_{ji} \geq 1$, for
each $\{i,j\}\in \vec{E}$. Then, we can derive the chain of inequalities
\[
\begin{split}
\Delta & =
(\alpha - 1) v_{ij} (V_i + \Gamma_{ij}^2 v_{ji} + \alpha S_{ij})
(V_i + \Gamma_{ij}^2 v_{ji} + S_{ij})
\\
& 
\quad
+
V_i + \Gamma_{ij}^2 v_{ji} + S_{ij}
-
\alpha(V_i + \Gamma_{ij}^2 v_{ji} + \alpha S_{ij})
\\
&
\geq
(\alpha - 1) v_{ij} (V_i  + \alpha S_{ij})
(V_i + \Gamma_{ij}^2 v_{ji} + S_{ij})
\\
&\quad
+
(\alpha - 1)
(V_i + \Gamma_{ij}^2 v_{ji} + S_{ij})
+
V_i + \Gamma_{ij}^2 v_{ji} + S_{ij}
\\
& 
\quad
-
\alpha(V_i + \Gamma_{ij}^2 v_{ji} + \alpha S_{ij})
\\
& 
=
(\alpha - 1) v_{ij} (V_i  + \alpha S_{ij})
(V_i + \Gamma_{ij}^2 v_{ji} + S_{ij})
\\
& 
\quad
-
\alpha(\alpha  - 1) S_{ij}
\\
&
\geq
(\alpha - 1) v_{ij} (V_i  + \alpha S_{ij})
(V_i + S_{ij})
\\
& 
\quad
+
(\alpha - 1) (V_i  + \alpha S_{ij})
-
\alpha(\alpha  - 1) S_{ij}
\\
&
=
(\alpha - 1) v_{ij} (V_i  + \alpha S_{ij})
(V_i + S_{ij})
+
(\alpha - 1) V_i 
\\
& > 0.
\end{split}
\]
\end{IEEEproof}

We are now ready to prove Theorem~\ref{th:gammaconv}.

\renewcommand{\thetheorem}{\ref{th:gammaconv}}
\begin{theorem}
  Assume that $f(\cdot)$ is convex decomposable.  The set of system of
  equations
\[
\gamma_{ij} = 
\frac{1}{1 - \sum_{u\in N(i)\setminus j} \Gamma_{ui}^2 \gamma_{ui}},
\quad\forall\ \{i,j\}\in\vec{E},
\]
has a solution $\gamma^*$ such that
\[
\0 < \gamma^* < v,\quad\forall\ v\in\Vscr.
\]
Moreover, $\gamma^*$ is the unique such solution.

If we initialize the min-sum algorithm so that $\gamma^{(0)} \leq v$,
for some $v\in\Vscr$, then $0 < \gamma^{(t)} < v,$ for all $t > 0$,
and
\[
\lim_{t\rightarrow\infty} 
\gamma^{(t)} = \gamma^*.
\]
\end{theorem}
\begin{IEEEproof}
  Pick some $v\in\Vscr$. Then, $F(v) < v$ from Part~(v) of
  Lemma~\ref{le:Fprop}. Thus, we have $F^t(v) \leq F^{t-1}(v)$, for
  all $t>0$, by monotonicity. (Here, $F^t(\cdot)$ denotes $t$
  applications of the operator $F(\cdot)$.) Then, the sequence $\{
  F^t(v)\}$ is a monotonically decreasing sequence, which by the
  positivity of $F(\cdot)$, is bounded below by zero. Hence, the limit
  $F^\infty(v)$ exists. By continuity, it must be a fixed point of
  $F(\cdot)$.

  Now, note that, by positivity, $\0 \leq F^\infty(v)$. Thus, by
  monotonicity, $F^t(\0) \leq F^{\infty}(v)$, for all $t>0$. Since $\0
  < F(\0) = \1$, we have $F^{t-1}(\0) \leq F^{t}(\0)$, for all $t>0$,
  and this sequence converges to a fixed point $F^\infty(\0) \leq
  F^\infty(v)$.

  We wish to show that $F^\infty(\0)=F^\infty(v)$. Assume
  otherwise. Define
\[
  \beta = \inf \{ \alpha\geq 1\ |\ v - \alpha(v-F^\infty(v)) \leq F^\infty(\0) \}.
\]
Since $F^\infty(v) < v$, the set in the above infimum is not
empty. Since $F^\infty(\0) \leq F^\infty(v)$ and $F^\infty(\0) \neq
F^\infty(v)$, we must have $\beta > 1$. Then, we have
\[
F^\infty(\0)  \geq v - \beta(v-F^\infty(v)).
\]
Applying $F(\cdot)$ and using Part~(iv) of Lemma~\ref{le:Fprop},
\[
\begin{split}
F^\infty(\0)  & \geq F(v - \beta(v-F^\infty(v))) \\
&
> \beta F^\infty(v) - (\beta - 1) v
\\
& 
=
v - \beta(v - F^\infty(v)).
\end{split}
\]
This contradicts the definition of $\beta$. Thus, we must have
$F^\infty(\0)=F^\infty(v)$.

Set $\gamma^*=F^\infty(\0)$. From the above argument, we have $\0 <
\gamma^*=F^\infty(v) < v$, for all $v\in \Vscr$.  Thus, $\gamma^*$
satisfies the conditions of the lemma.

Assume there is some other fixed point $\gamma'$ satisfying the
conditions of the lemma. Positivity implies $\gamma' > 0$. Then, since
$\0 < \gamma' < v$ for some $v\in\Vscr$, by repeatedly applying
$F(\cdot)$, we have
\[
F^{t}(\0) \leq \gamma' \leq
F^{t}(v),
\] 
for all $t > 0$. Taking a limit as $t\rightarrow\infty$, it is clear that
$\gamma'=\gamma*$.

It remains to prove the final statement of the lemma. Consider
$\gamma^{(0)}$, with $\gamma^{(0)} \leq v$, for some $v\in\Vscr$.  Note that $\0 <
F(\gamma) \leq F(v) < v$. Then, 
\[
\0 < F^{t}(\0) \leq \gamma^{(t+1)}=F^{t+1}(\gamma^{(0)}) \leq F^{t+1}(v) < v.
\]
for all $t > 0$. Taking limits, 
\[
\lim_{t\rightarrow\infty} 
\gamma^{(t)} = \gamma^*.
\]
\end{IEEEproof}

\section{Proof of Lemmas~\ref{le:nuwalksum} and
  \ref{le:nuexact}}\label{ap:walksum}

For the balance of this section, we assume that $f(\cdot)$ admits a
convex decomposition.

In order to prove Lemma~\ref{le:nuwalksum}, we first fix an arbitrary
vertex $r$, and consider an infinite computation tree rooted at a
vertex $\tilde{r}$ corresponding to $r$. Such a tree is constructed in
an iterative process, first starting with a single vertex
$\tilde{r}$. As each step, vertices are added to leaves on the tree
corresponding to the neighbors of the leaf in the original graph other
than its parent. Hence, the tree's vertices consist of replicas of
vertices in the original graph, and the local structure around each
vertex is the same as that in the original graph.  We can extend both
functions $\rho(\cdot)$ and $\nu(\cdot)$ to walks on the computation
tree by defining weights on edges in the computation tree according to
the weights of the corresponding edges in the original graph. We will
use the tilde symbol to distinguish vertices and subsets of the
computation tree from those in the underlying graph.

We begin with a lemma.

\begin{lemma}\label{le:srminus}
  Given connected vertices $\tilde{i},\tilde{j}$ in the computation tree,
  with labels $i$,$j$, respectively, let
  $\tilde{\Wscr}_{\tilde{i}\rightarrow \tilde{i}\setminus \tilde{j}}$
  be the set of walks starting at $\tilde{i}$ and returning to
  $\tilde{i}$ but never crossing the edge
  $(\tilde{i},\tilde{j})$. Then,
\[
\rho(\tilde{\Wscr}_{\tilde{i}\rightarrow \tilde{i}\setminus \tilde{j}}) = \gamma^*_{ij}.
\]
\end{lemma}
\begin{IEEEproof}
  First, note that walks in $\tilde{\Wscr}_{\tilde{i}\rightarrow
    \tilde{i}\setminus \tilde{j}}$ can be mapped to disjoint walks on
  the original graph. Hence, by walk-summability, the infinite sum
\[
\sum_{\tilde{w}\in \tilde{\Wscr}_{\tilde{i}\rightarrow \tilde{i}\setminus \tilde{j}}}
\rho(\tilde{w})
\]
converges absolutely.

Now, define the set $\tilde{\Wscr}^d_{\tilde{i}\rightarrow
  \tilde{i}\setminus \tilde{j}}$ to be the set of walks in
$\tilde{\Wscr}_{\tilde{i}\rightarrow \tilde{i}\setminus \tilde{j}}$
that travel at most a distance $d$ away from $\tilde{i}$ in the
computation tree.  A walk $\tilde{w}\in
\tilde{\Wscr}^d_{\tilde{i}\rightarrow \tilde{i}\setminus \tilde{j}}$
can be decomposed into a series of traversals to neighbors
$\tilde{u}\in N(\tilde{i})\setminus \tilde{j}$, self-returning walks
from $\tilde{u}$ to $\tilde{u}$ that do not cross
$(\tilde{u},\tilde{i})$ and travel at most distance $d-1$ from
$\tilde{u}$, and then returns to $\tilde{i}$. Letting $t$ index the
total number of such traversals, we have the expression
\[
\rho(\tilde{\Wscr}^d_{\tilde{i}\rightarrow \tilde{i}\setminus \tilde{j}})
=
\sum_{t=0}^\infty
\left(
\sum_{\tilde{u}\in N(\tilde{i})\setminus \tilde{j}}
R_{\tilde{u} \tilde{i}}^2 
\rho(\tilde{\Wscr}^{d-1}_{\tilde{u}\rightarrow \tilde{u}\setminus \tilde{i}})
\right)^t.
\]
By walk-summability, this infinite sum must converge. Thus,
\[
\rho(\tilde{\Wscr}^d_{\tilde{i}\rightarrow \tilde{i}\setminus \tilde{j}})
=
\frac{1}
{1 - 
\sum_{\tilde{u}\in N(\tilde{i})\setminus \tilde{j}}
R_{\tilde{u} \tilde{i}}^2 
\rho(\tilde{\Wscr}^{d-1}_{\tilde{u}\rightarrow \tilde{u}\setminus \tilde{i}})
}.
\]
By the symmetry of the computation tree, the quantity
$\rho(\tilde{\Wscr}^d_{\tilde{i}\rightarrow \tilde{i}\setminus
  \tilde{j}})$ depends only on the labels of $\tilde{i}$ and
$\tilde{j}$ in the original graph. Set $\gamma^{(0)}_{ij} =
\rho(\tilde{\Wscr}^0_{\tilde{i}\rightarrow \tilde{i}\setminus
  \tilde{j}}) = 1$ and $\gamma^{(d)}_{ij} =
\rho(\tilde{\Wscr}^d_{\tilde{i}\rightarrow \tilde{i}\setminus
  \tilde{j}})$, for each $\{i,j\}\in\vec{E}$ and integer $d > 0$. Then, we have 
\[
\gamma^{(d)}_{ij}
=
\frac{1}
{1 - 
\sum_{u\in N(i)\setminus j}
R_{ui}^2 
\gamma^{(d-1)}_{ui}}.
\]
By Theorem~\ref{th:gammaconv}, we have
\[
\lim_{d\rightarrow\infty} \gamma^{(d)}_{ij} = \gamma^*_{ij}.
\]
Then, since $\tilde{\Wscr}^d_{\tilde{i}\rightarrow \tilde{i}\setminus
  \tilde{j}} \subset \tilde{\Wscr}^{d+1}_{\tilde{i}\rightarrow \tilde{i}\setminus
  \tilde{j}}$, and
\[
\tilde{\Wscr}_{\tilde{i}\rightarrow \tilde{i}\setminus
  \tilde{j}}
=
\bigcup_{d=0}^\infty
\tilde{\Wscr}^d_{\tilde{i}\rightarrow \tilde{i}\setminus
  \tilde{j}},
\]
we have
\[
\rho(\tilde{\Wscr}_{\tilde{i}\rightarrow \tilde{i}\setminus
  \tilde{j}})
=
\lim_{d\rightarrow\infty} \rho(\tilde{\Wscr}^d_{\tilde{i}\rightarrow \tilde{i}\setminus
  \tilde{j}}) = \gamma^*_{ij}.
\]
\end{IEEEproof}

We call a walk on the computation tree a {\it shortest-path} walk if
it is the unique shortest path between its endpoints. Given a
shortest-path walk $\tilde{p}$ define $\tilde{\Wscr}_{\tilde{p}}$ to
be the set of all walks of the form
\[
\{\tilde{p_0},\tilde{w}^0,\tilde{p_1},\tilde{w}^1,\ldots,\tilde{w}^{|\tilde{p}|-1},\tilde{p}_{|\tilde{p}|}\},
\]
where $\tilde{w}^i\in
\tilde{\Wscr}_{\tilde{p}_i\rightarrow\tilde{p}_i\setminus\tilde{p}_{i+1}}$,
for $0 \leq i < |\tilde{p}|$. Intuitively, these walks proceed along
the path $p$, but at each point $\tilde{p}_i$, they may also take a
self-returning walk from vertex $\tilde{p}_i$ to vertex $\tilde{p}_i$
that does not cross the edge $(\tilde{p}_i,\tilde{p}_{i+1})$.

\begin{lemma}\label{le:tildep}
Given a shortest-path walk $\tilde{p}$,
\[
\rho(\tilde{\Wscr}_{\tilde{p}}) = \nu(\tilde{p}).
\]
\end{lemma}
\begin{IEEEproof}
\[
\begin{split}
\rho(\tilde{\Wscr}_{\tilde{p}}) 
& = 
\sum_{\tilde{w}^0\in
\tilde{\Wscr}_{\tilde{p}_0\rightarrow\tilde{p}_0\setminus\tilde{p}_{1}}}
\cdots
\sum_{\tilde{w}^{|\tilde{p}|-1}\in
\tilde{\Wscr}_{\tilde{p}_0\rightarrow\tilde{p}_{|\tilde{p}|-1}\setminus\tilde{p}_{|\tilde{p}|}}}
\\
&\qquad\qquad\qquad
\rho(\{\tilde{p_0},\tilde{w}^0,\tilde{p_1},\tilde{w}^1,\ldots,\tilde{w}^{|\tilde{p}|-1},\tilde{p}_{|\tilde{p}|}\})
\\
& =
\rho(\tilde{p})
\prod_{i=0}^{|\tilde{p}|-1}
\rho(\tilde{\Wscr}_{\tilde{p}_i\rightarrow\tilde{p}_{i}\setminus\tilde{p}_{i+1}})
\\
& =
\nu(\tilde{p}).
\end{split}
\]
\end{IEEEproof}

We are now ready to prove Lemma~\ref{le:nuwalksum}.

\renewcommand{\thelemma}{\ref{le:nuwalksum}}
\begin{lemma}
  Assume that $f(\cdot)$ is convex decomposable.
  For each $w\in\Wscr^{nb}$, there exists a set of walks $\Wscr_{w}$,
  all terminating at the same vertex as $w$, such that
\[
\nu(w) = \rho(\Wscr_{w}).
\]
Further, if $w'\in\Wscr^{nb}$ and $w'\neq w$, then $\Wscr_w$ and
$\Wscr_{w'}$ are disjoint.
\end{lemma}
\begin{IEEEproof}
  Take a vertex $i$ in the original graph. Given a walk from $i$ to
  $r$ in the original graph, there is a unique corresponding walk from
  a replica of $i$ to $\tilde{r}$ in the computation tree.  Also
  notice that non-backtracking walks in the original graph that
  terminate at $r$ correspond uniquely to shortest-path walks in the
  computation tree that terminate at $\tilde{r}$.

Now, assume that $w\in \Wscr^{nb}$ terminates at $r$. Let $\tilde{p}$
be the corresponding shortest-path walk in the computation tree, and
consider the set $\tilde{W}_{\tilde{p}}$. We will define $\Wscr_w$ to
be the set of walks in the original graph corresponding to
$\tilde{W}_{\tilde{p}}$. From Lemma~\ref{le:tildep},
\[
\nu(w) = \nu(\tilde{p}) = \rho(\tilde{W}_{\tilde{p}}) = \rho(\Wscr_w).
\]

Now, consider another walk $w'\in\Wscr^{nb}$, $w'\neq w$, that also
terminates at $r$. We would like to show that $\Wscr_w$ and
$\Wscr_{w'}$ are disjoint. Let $\tilde{p}'$ be the shortest-path walk
corresponding to $w'$. Equivalently, we can show
$\tilde{W}_{\tilde{p}}$ and $\tilde{W}_{\tilde{p}'}$ are

disjoint. Assume there is some walk $\tilde{u}\in
\tilde{W}_{\tilde{p}} \cap \tilde{W}_{\tilde{p}'}$. Then, both
$\tilde{p}$ and $\tilde{p}'$ must be the shortest-path from the origin
of $\tilde{u}$ to $\tilde{r}$. Since shortest-paths between a pair of
vertices on the computation tree are unique, we must have
$\tilde{p}=\tilde{p}'$ and this $w=w'$, which is a contradiction.

Note that we only considered non-backtracking walks terminating at a
fixed vertex $r$. However, our choice or $r$ was arbitrary hence we can
repeat the construction for each $r\in V$. Moreover, if $w$ and $w'$
terminate at different vertices $r$ and $r'$, respectively, the sets
$\Wscr_w$ and $\Wscr_{w'}$ will contain only walks that terminate at
$r$ and $r'$, respectively, thus they will be disjoint.
\end{IEEEproof}

Using similar arguments as above, we can prove Lemma~\ref{le:nuexact}.

\renewcommand{\thelemma}{\ref{le:nuexact}}
\begin{lemma}
  Assume that $f(\cdot)$ is convex decomposable.
  If we define
  $\Wscr^{nb}_{i\rightarrow r}$ to be the set of all non-backtracking
  walks from vertex $i$ to vertex $r$, we have
\[
\rho(\Wscr_{i\rightarrow r}) = \frac{\nu(\Wscr^{nb}_{i \rightarrow r})}
{1 - \sum_{u\in N(r)} R_{ur}^2 \gamma^*_{ur}}.
\]
\end{lemma}
\begin{IEEEproof}
  Consider a walk $w\in\Wscr_{i \rightarrow r}$, and let $\tilde{w}$
  be the unique corresponding walk in the computation tree terminating
  at $\tilde{r}$. Let $\tilde{p}$ be the unique shortest-path walk
  corresponding to $\tilde{w}$. Note that $\tilde{p}$ will originate
  at a replica of $i$, and end at $\tilde{r}$. Thus, $\tilde{p}$
  uniquely corresponds to a non-backtracking walk $w'\in\Wscr^{nb}_{i
    \rightarrow r}$.

Now, $\tilde{w}$ can be uniquely decomposed
  according to
\[
\{\tilde{p_0},\tilde{w}^0,\tilde{p_1},\tilde{w}^1,\ldots,\tilde{w}^{|\tilde{p}|-1},\tilde{p}_{|\tilde{p}|},\tilde{v}\},
\]
where $\tilde{w}^i\in
\tilde{\Wscr}_{\tilde{p}_i\rightarrow\tilde{p}_i\setminus\tilde{p}_{i+1}}$,
for $0 \leq i < |\tilde{p}|$, and $\tilde{v}$ is a self-returning walk
from $\tilde{r}$ to $\tilde{r}$. Applying Lemma~\ref{le:tildep}, we have
\[
\rho(\Wscr_{i\rightarrow r}) = \nu(\Wscr^{nb}_{i \rightarrow r})
\rho(\tilde{\Wscr}_{\tilde{r}\rightarrow\tilde{r}}),
\]
where $\tilde{\Wscr}_{\tilde{r}\rightarrow\tilde{r}}$ is the set of
self-returning walks from $\tilde{r}$ to $\tilde{r}$.

However, a walk
$\tilde{v}\in\tilde{\Wscr}_{\tilde{r}\rightarrow\tilde{r}}$ can be
uniquely decomposed into a series of traversals to neighbors
$\tilde{u}\in N(\tilde{r})$, self-returning walks from $\tilde{u}$ to
$\tilde{u}$ that do not cross $(\tilde{u},\tilde{r})$, and then
returns to $\tilde{i}$. Letting $t$ index the total number of such
traversals, we have the expression
\[
\begin{split}
\rho(\tilde{\Wscr}_{\tilde{r}\rightarrow\tilde{r}})
& =
\sum_{t=0}^\infty
\left(
\sum_{\tilde{u}\in N(\tilde{r})}
R_{\tilde{u} \tilde{r}}^2 
\rho(\tilde{\Wscr}_{\tilde{u}\rightarrow \tilde{u}\setminus \tilde{r}})
\right)^t.
\end{split}
\]
From Lemma~\ref{le:srminus},
\[
\rho(\tilde{\Wscr}_{\tilde{u}\rightarrow \tilde{u}\setminus \tilde{r}})
= \gamma^*_{ur}.
\]
Thus,
\[
\begin{split}
\rho(\Wscr_{i\rightarrow r})
& =
\nu(\Wscr^{nb}_{i \rightarrow r})
\sum_{t=0}^\infty
\left(
\sum_{u\in N(r)}
R_{u r}^2 
\gamma^*_{ur}
\right)^t
\\
& =
\frac{
\nu(\Wscr^{nb}_{i \rightarrow r})
}
{
1 -
\sum_{u\in N(r)}
R_{u r}^2 
\gamma^*_{ur}
}.
\end{split}
\]
\end{IEEEproof}

\section*{Acknowledgment}

The first author wishes to thank Jason Johnson for a helpful
discussion.  The first author was supported by a Benchmark Stanford
Graduate Fellowship.  This material is based upon work supported by 
the National Science Foundation under Grant No. IIS-0428868.

\ifCLASSOPTIONcaptionsoff
  \newpage
\fi


\vspace{2.0in}

\begin{IEEEbiography}{Ciamac C. Moallemi}
  Ciamac C. Moallemi is an Assistant Professor at the Graduate School
  of Business of Columbia University, where he has been since 2007. He
  received SB degrees in Electrical Engineering \& Computer Science
  and in Mathematics from the Massachusetts Institute of Technology
  (1996). He studied at the University of Cambridge, where he earned a
  Certificate of Advanced Study in Mathematics, with distinction
  (1997). He received a PhD in Electrical Engineering from Stanford
  University (2007). He is a member of the IEEE and INFORMS. He is the
  receipient of a British Marshall Scholarship (1996) and a Benchmark
  Stanford Graduate Fellowship (2003).
\end{IEEEbiography}

\begin{IEEEbiography}{Benjamin Van Roy}
  Benjamin Van Roy is an Associate Professor of Management Science and
  Engineering, Electrical Engineering, and, by courtesy, Computer
  Science, at Stanford University. He has held visiting positions as
  the Wolfgang and Helga Gaul Visiting Professor at the University of
  Karlsruhe and as the Chin Sophonpanich Foundation Professor of
  Banking and Finance at Chulalongkorn University.  He received the SB
  (1993) in Computer Science and Engineering and the SM (1995) and PhD
  (1998) in Electrical Engineering and Computer Science, all from
  MIT. He is a member of INFORMS and IEEE. He has served on the
  editorial boards of Discrete Event Dynamic Systems, Machine
  Learning, Mathematics of Operations Research, and Operations
  Research.  He has been a recipient of the MIT George C. Newton
  Undergraduate Laboratory Project Award (1993), the MIT Morris
  J. Levin Memorial Master's Thesis Award (1995), the MIT George
  M. Sprowls Doctoral Dissertation Award (1998), the NSF CAREER Award
  (2000), and the Stanford Tau Beta Pi Award for Excellence in
  Undergraduate Teaching (2003). He has been a Frederick E. Terman
  Fellow and a David Morgenthaler II Faculty Scholar.
\end{IEEEbiography}

\end{document}